\input{aipcheck.tex}

\documentclass[
    ,final            
  ]
  {aipproc}

\layoutstyle{8x11single}

\begin{document}

\title{ANTARES Deep Sea Neutrino Telescope Results}


\author{Salvatore Mangano, on behalf of the ANTARES Collaboration}{
  address={IFIC - Instituto de F\'isica Corpuscular, Edificio Institutos de Investigati\'on, 46071 Valencia, Spain}
}



\begin{abstract}
The ANTARES experiment is currently the largest underwater neutrino
telescope in the Northern Hemisphere. It is taking high quality data since 2007. 
Its main scientific goal is to search for high
energy neutrinos that are expected from the acceleration of cosmic rays
from astrophysical sources. This contribution reviews the status of the detector
and presents several analyses carried out on atmospheric muons and
neutrinos. For example it shows the results from 
the measurement of atmospheric muon neutrino spectrum and 
of atmospheric neutrino oscillation parameters as well as 
searches for neutrinos from steady cosmic point-like sources,
for neutrinos from gamma ray bursts and 
for relativistic magnetic monopoles.

\end{abstract}

\maketitle


\section{Neutrino Astronomy}
Neutrino astronomy has a unique opportunity to observe processes that are inaccessible 
to optical telescopes or cosmic ray observatories.  
The advantage of neutrinos with respect to cosmic particles like  photons and protons 
is that they can travel over cosmological distances without being 
absorbed or deflected by magnetic fields.

The existence of high energy cosmic rays is known since over 100 years, 
but their astrophysical origin and their acceleration to
the high energies is still unclear.
The observation of cosmic rays is a strong argument 
for the existence of high energy neutrinos from the cosmos. 
The cosmic neutrinos are expected to be emitted along with gamma-rays by
astrophysical sources in processes
involving the interaction of accelerated hadrons with ambient matter or dense photon fields.
The subsequent production and decay of pions produce high-energy neutrinos and  photons.

The weak interaction of neutrinos with matter makes the detection challenging.
A cost effective way to detect high energy neutrinos is to use target 
material found in nature, like water and ice.
The detector material has to be equipped with a three dimensional array of 
light sensors, so that muon neutrinos are
identified by the muons that are produced in charged current
interactions. These muons are detected by measuring the
Cerenkov light which they emit when charged particles move faster than 
the speed of light in the detector material.
The knowledge of the timing of the Cerenkov 
light recorded by 
the light sensors allows to reconstruct the trajectory of the muon 
and so to infer the arrival direction of the incident neutrino.
This technique is used in
large-scale Cerenkov detectors like IceCube~\citet{icecube} and ANTARES~\citet{Ant1} which are 
currently looking for high-energy
($>$TeV) cosmic neutrinos.

\section{ANTARES Neutrino Telescope}
The ANTARES (\textbf{A}stronomy with a \textbf{N}eutrino \textbf{T}elescope and \textbf{A}byss environmental \textbf{RES}earch) detector is taking data since the first lines were deployed in 
2006. It is located in the
Mediterranean Sea, \mbox{40 km} off the French coast at \mbox{$42^{\circ} 50'$N, $6^{\circ} 10'$E}.
The detector consists of twelve vertical
lines equipped with 885 photomultipliers (PMTs) in total, 
installed at a depth of about 2.5~km. 
The distance between adjacent lines is of the order of about 70 m.
Each line is equipped with up to 25 triplets of PMTs 
spaced vertically by 14.5 m. The
PMTs are oriented with their axis 
pointing downwards at $45^{\circ}$ from the vertical. 
The instrumented detector volume is about 0.02 $\textrm{km}^3$.
The design of ANTARES is optimized for the detection of upward going muons 
produced by neutrinos which have traversed the Earth, 
in order to limit the background from downward going atmospheric muons. 
The instantaneous field of view is half of the sky for neutrino 
energies between $10\ \mathrm{GeV}$ and $100\  \mathrm{TeV}$, 
due to selection of upgoing events and the size of the detector.
Further details on the detector can be 
found elsewhere~\cite{Ant1}.

In this proceeding there is not 
enough room to discuss all
topics which were illustrated at Particle Physics and Cosmology workshop, 
such as the atmospheric muon~\cite{4gev}, diffuse neutrino fluxes~\cite{diffusepaper}, neutrinos from dark matter from the Sun~\cite{Dmatter}, the time calibration system~\cite{timecalib} and the acoustic neutrino detection system~\cite{acoustic}, for which
the reader is referred to elsewhere. A short description of some interesting 
measurements and searches using ANTARES data are presented in the following.


\subsection{Measurement of Atmospheric Muon Neutrino Spectrum}
\begin{figure}   
  \setlength{\unitlength}{1cm}      
   \centering
   \begin{picture}(18.0,6.0)
     \put(0.4, 0.0){\includegraphics[height=.28\textheight]{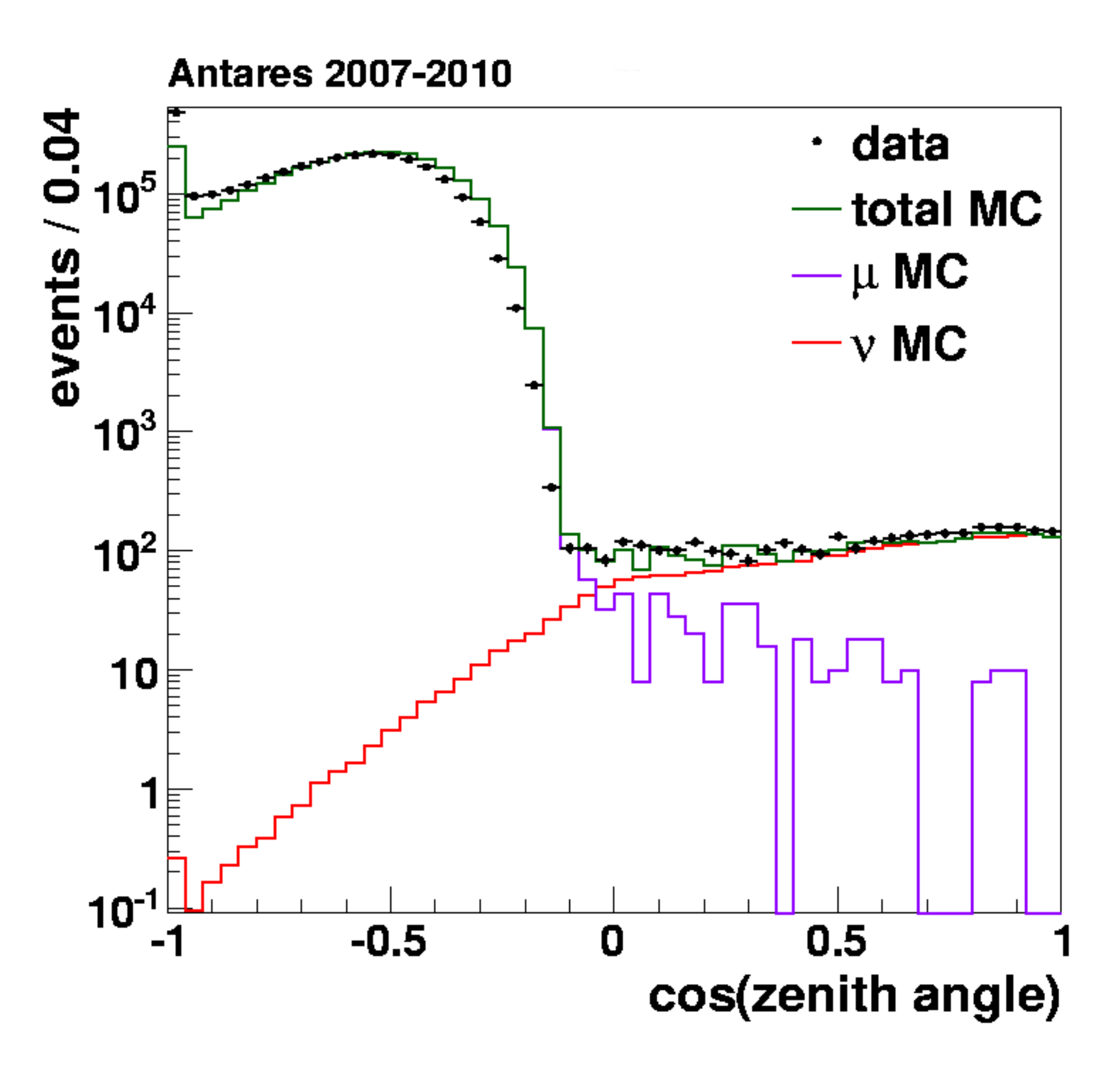}}
     \put(7.2, 0.0){\includegraphics[height=.265\textheight]{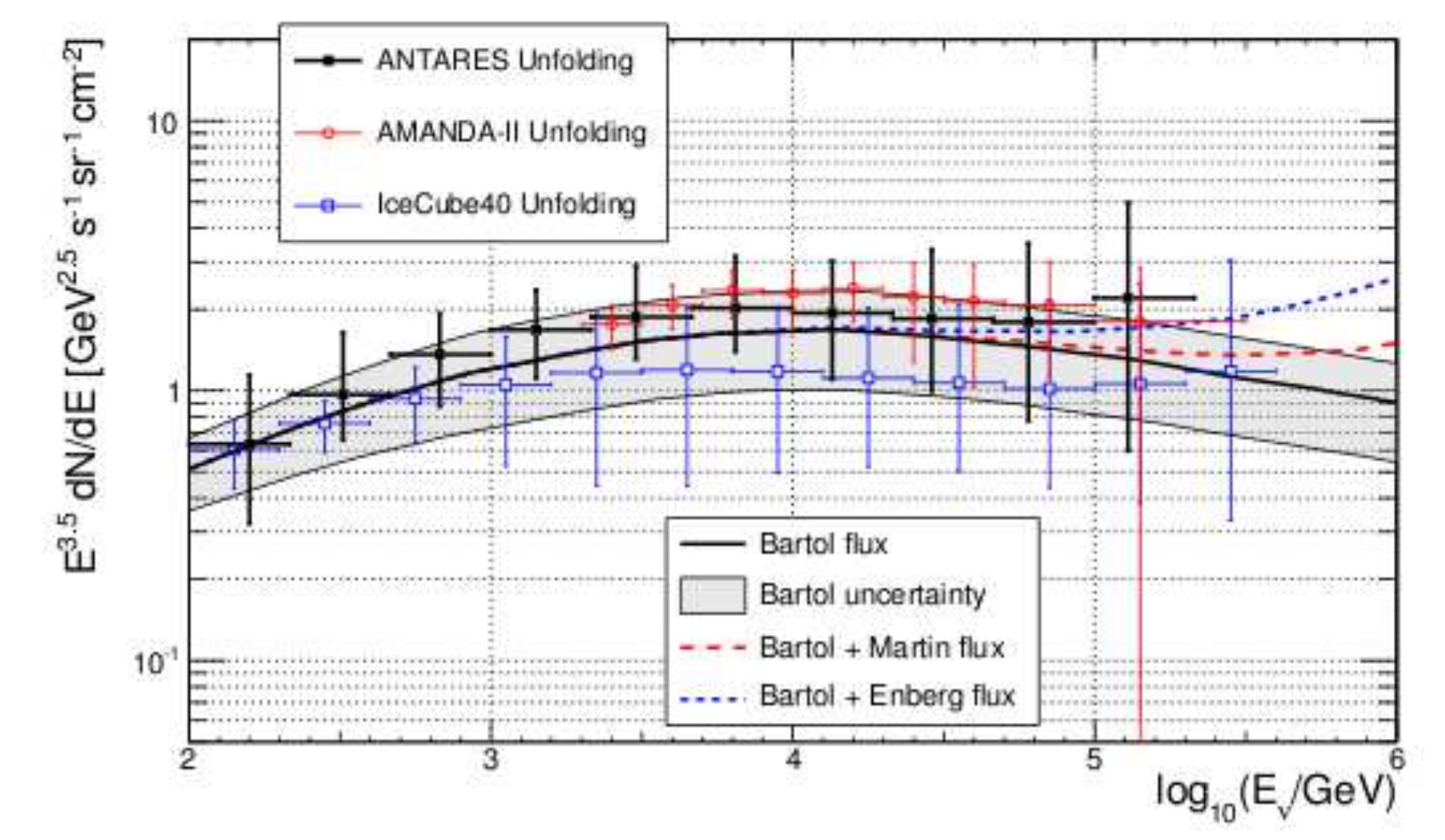}}
     \end{picture}
  \caption{\textit{Left: The reconstructed  zenith angle data distribution of selected events compared to the Monte Carlo distribution for atmospheric neutrino and muon background. Right: The zenith-averaged atmospheric neutrino energy spectrum. The ANTARES result is shown together with the results from AMANDA-II~\cite{amandaflux} and IceCube40~\cite{icecubeflux}. The solid black line represents the conventional flux prediction from Bartol group and the shaded area represents its uncertainty~\cite{barr}. The dashed red and dotted blue lines include two prompt neutrino production models from~\cite{Martin} and~\cite{Enberg}, respectively. }}
\label{fig:zenithneutrinoflux}
\end{figure}

Even if the primary aim of ANTARES is the detection of high energy
cosmic neutrinos,  the detector measures mainly downward going atmospheric muons and 
upward going atmospheric muon neutrinos.
The atmospheric muons are produced in the upper
atmosphere by the interaction of cosmic rays and can 
reach the apparatus despite the
shielding provided by 2 km of water. 
Atmospheric neutrinos produced also in the atmospheric cascades as the above mentioned atmospheric 
muons, can travel through
the Earth and interact in the vicinity of the detector, producing the upward going 
atmospheric neutrinos.
Figure \ref{fig:zenithneutrinoflux} left shows a 
comparison of the zenith 
angle distribution between data and 
Monte Carlo simulation. It can be seen that 
the flux of atmospheric muons 
is several order of magnitude larger than that of atmospheric neutrinos
and that there is a good agreement between data and the Monte Carlo simulation.

The measurement of the atmospheric muon neutrino spectrum 
has been performed using 2008-2011 data for 
a total equivalent live time of 855 days.
A determination of the neutrino energy is needed for such a measurement. 
In the analysis two energy estimators were used.
The first one is based on the muon energy loss along its trajectory and the second one
relies on a maximum likelihood method attempting
to maximize the agreement between the observed and expected amount of light.  
To reconstruct the energy spectrum an unfolding procedure is used for both methods and
the results are represented in Figure~\ref{fig:zenithneutrinoflux} right.  
Within the errors the ANTARES results are compatible with the spectra measured 
in the Antarctic neutrino telescopes.
Also shown is the distinction between neutrinos produced by the decay of pions 
and kaons up to about 100 TeV, the 
so-called conventional neutrinos and
the neutrinos produced by the decay of charmed mesons, 
the so-called prompt neutrinos~\cite{Martin,Enberg}. 
The energy dependence of the prompt neutrino flux are poorly constrained. Its precise
features are sensitive to hadronic interaction models and 
is less steep than the conventional neutrino flux. The highest energy region of the 
atmospheric neutrino spectrum has been used to put a constraint on the diffuse of cosmic neutrinos~\cite{diffusepaper}.
Such a diffuse flux would reflect the existence of a cumulative neutrino flux from a bulk of 
unresolved astrophysical sources.

\subsection{Measurement of Atmospheric Muon Neutrino Oscillations}
At lowest neutrino energy ANTARES is sensitive to neutrino oscillation parameters 
through the disappearance of atmospheric muon neutrinos~\cite{oscillation}.
Neutrino oscillation are commonly described
in terms of L/E, where L the oscillation path length 
and E is the neutrino energy. 
For upward going neutrinos crossing the Earth  
the travel distance L is translated to $D \cos\theta$ where
$D$ is the Earth diameter and $\theta$ the zenith angle.
Within the two-flavor approximation, the $\nu_{\mu}$
survival probability can be written as

$$P(\nu_{\mu} \rightarrow \nu_{\mu}) = 1 - \sin^2 2\theta_{23} \cdot \sin^2(1.27 \triangle m^2_{23}\frac{L}{E_{\nu}}) = 1 - \sin^2 2\theta_{23} \cdot
 \sin^2(16200 \triangle m^2_{23}\frac{\cos \theta}{E_{\nu}}),$$

\noindent where $\theta_{23}$ is the mixing angle and $\triangle m^2_{23}$ is the squared
mass difference of the mass eigenstates (with $L$ in km, $E _{\nu}$ in GeV and $\triangle m^2_{23}$ in eV$^2$). The survival probability $P$ depends
only on the two oscillation parameters, $\sin^2 2\theta_{23}$ and $\triangle m^2_{23}$,
which determine the behavior for the atmospheric neutrino oscillations.

Taking the recent results from the MINOS experiment~\cite{minos}, the first minimum
in the muon neutrino survival probability 
$(P(\nu_{\mu} \rightarrow \nu_{\mu})=0)$ occurs for vertical upward going neutrinos
at about 24 GeV. 
Muons induced by a 24 GeV neutrino travel in average around 120 m in sea water.
The detector has PMTs spaced vertically by 14.5 m so that this energy range 
can be reached for events detected on one single line. 

\begin{figure}   
  \setlength{\unitlength}{1cm}      
   \centering
   \begin{picture}(18.0,6.0)
     \put(-0.0, 0.0){\includegraphics[height=.28\textheight]{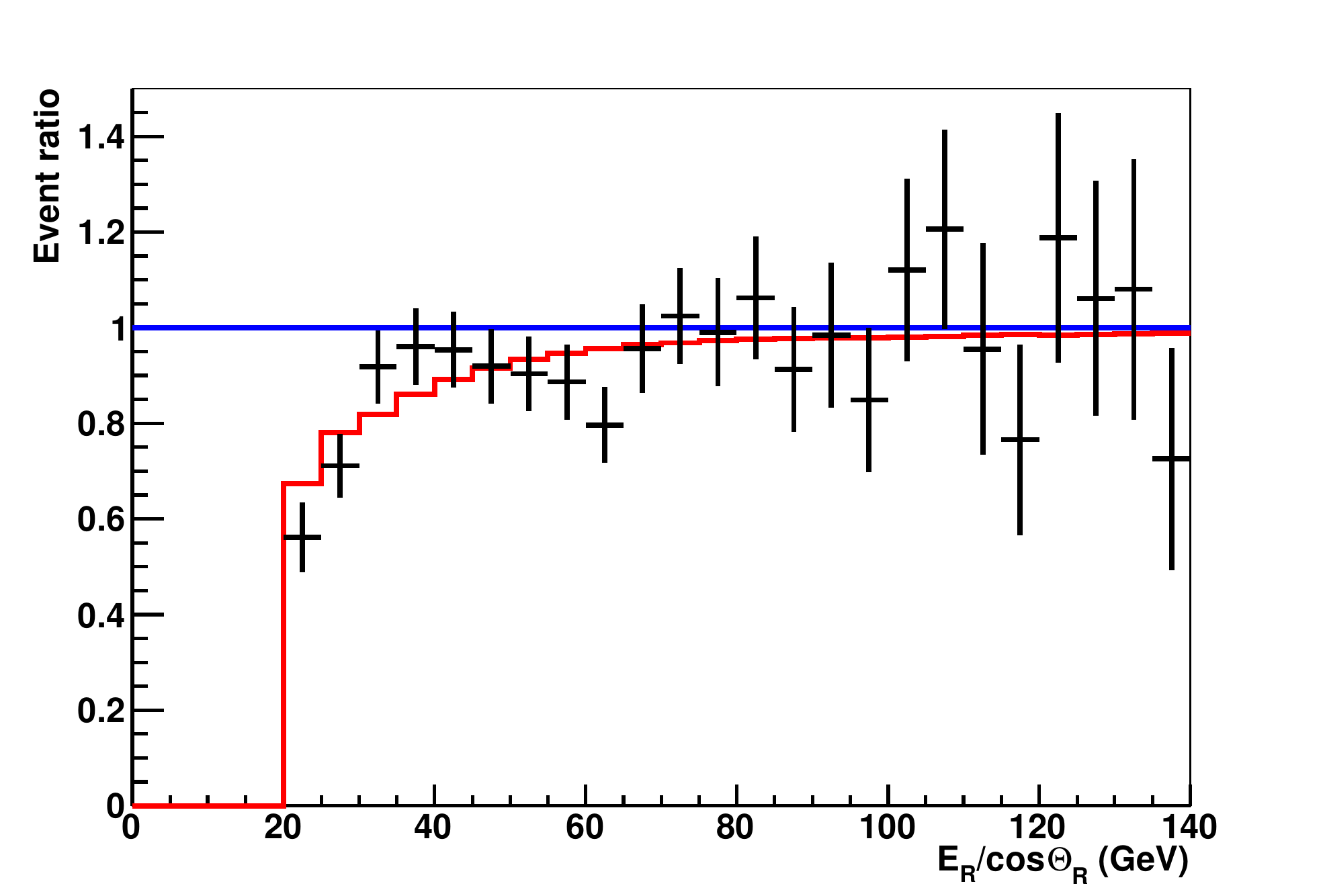}}
     \put(8.8, 0.0){\includegraphics[height=.28\textheight]{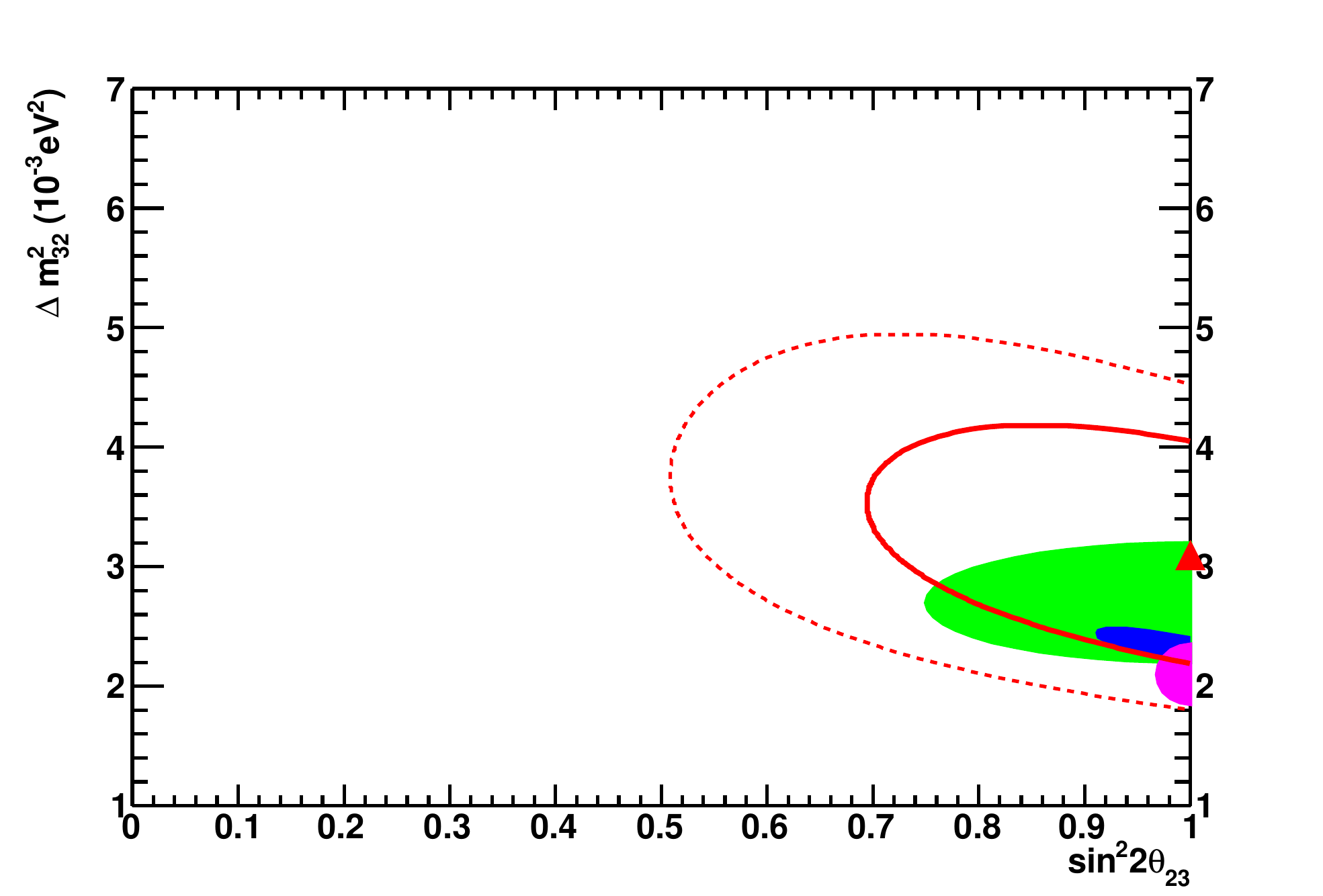}}
     \end{picture}
  \caption{\textit{Left: The fraction of events as a function of the $E_R/cos\theta_R$ distribution. Black crosses are data with statistical uncertainties, the blue straight line shows the non-oscillation hypothesis and the red line shows the result of the best fit. Right: 68\% and 90\% C.L. contours (solid and dashed red lines) of the neutrino oscillation parameters as derived from the fit of the $E_R/cos\theta_R$ distribution. The best fit point is indicated by the triangle. For comparison the solid filled regions show results at 68\% C.L. from K2K (green), MINOS (blue) and Super-Kamiokande (magenta).}}
\label{fig:oscillation}
\end{figure}

The reconstructed flight 
path through the Earth is reconstructed through zenith angle $\theta_R$, 
which is estimated from a muon track fit~\cite{fastalgo}. Whereas
the neutrino energy $E_R$ is estimated from 
the observed muon range in the detector.
Figure \ref{fig:oscillation} left 
shows event rate of the the measured variable 
$E_R/\cos\theta_R$ for a data sample from 2007 to 2010 with 
a total live time of 863 days.
Neutrino oscillations cause a clear 
event suppression for $E_R/\cos\theta_R < 60$ GeV with a clean sample of 
atmospheric neutrinos 
with energies as low as 20 GeV.
The parameters of the atmospheric neutrino oscillations 
are extracted by fitting the event rate as a function $E_R/\cos\theta_R$ and
is plotted as a red curve in Figure \ref{fig:oscillation} left with
values $\triangle m^2_{23}=3.1\cdot 10^{-3}~\mathrm{eV}^2$ and $\sin^2 2\theta_{23} = 1$.

This measurement is converted into limits of the oscillation parameters and
is shown in Figure \ref{fig:oscillation} right. 
If maximum mixing is imposed ($\sin^2 2\theta_{23} = 1$) the value is 
$\triangle m^2_{23}=(3.1 \pm 0.9) \cdot 10^{-3}~\mathrm{eV}^2$.
This measurement is in good agreement with the world average value.
Although the results are not competitive with dedicated experiments,
the ANTARES detector demonstrates the capability to measure
atmospheric neutrino oscillation parameters 
and to detect and measure energies as low as 20 GeV. It was the first time that a high energy neutrino
telescope was used to measure the atmospheric neutrino oscillation parameters.

\begin{figure}   
  \setlength{\unitlength}{1cm}      
   \centering
   \begin{picture}(18.0,6.0)
     \put(1.0, 0.0){\includegraphics[height=.28\textheight]{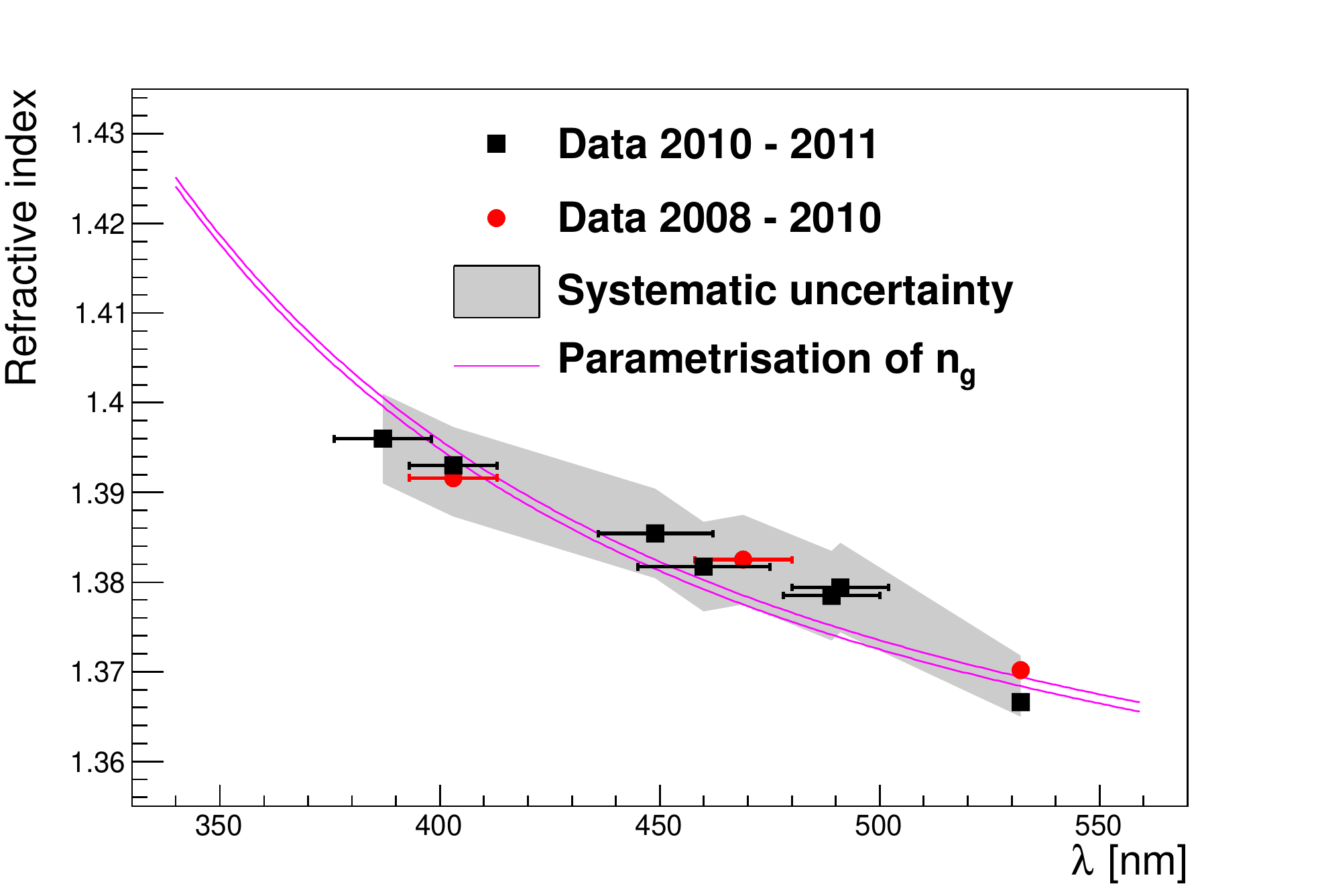}}
     \put(11.3, 0.0){\includegraphics[height=.28\textheight]{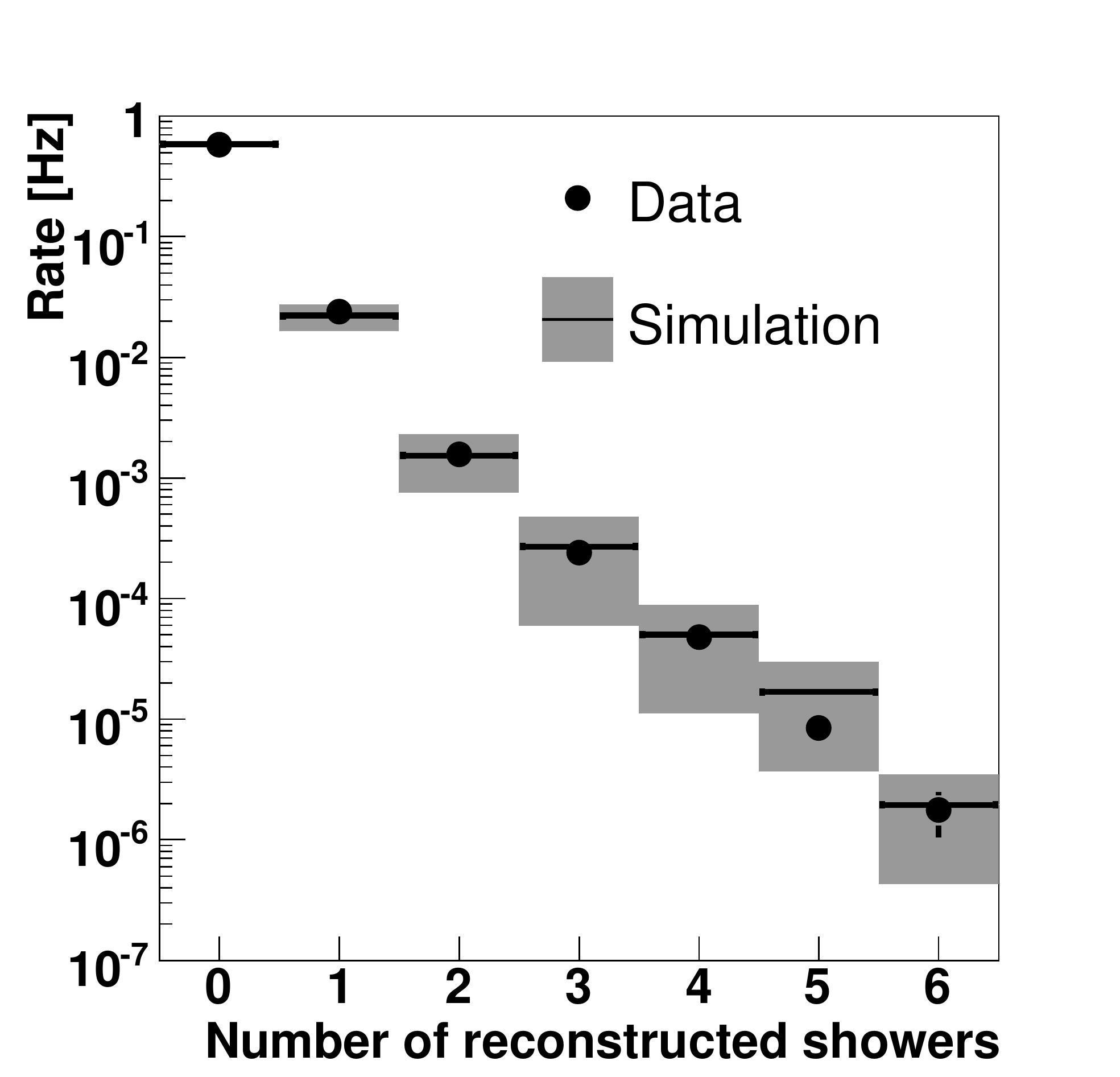}}
     \end{picture}
  \caption{\textit{Left: Index of refraction corresponding to the group velocity of light as a function of the wavelength. The gray band shows the systematic uncertainty. The two solid lines correspond to a parametrization of the index of refraction evaluated at a pressure of 200 atm (lower line) and 240 atm (upper line). Right: Muon event rate as a function of the shower multiplicity for data (points) and the Corsika simulation (line). The systematic error for the simulation is given by the vertical size of the gray bands.}}
\label{fig:showers}
\end{figure}

\subsection{Measurement of Velocity of Light in Water}
The correct understanding of the velocity of light in
the water at the detector site is crucial to reach the optimal performance of the detector.
It is well known that charged particles crossing sea water induce the emission of Cerenkov
light whenever the condition $\beta > 1/n_p$ is fulfilled, where
$\beta$ is the speed of the particle relative to the
speed of light in vacuum and $n_p$ is the phase refractive index. 
The Cerenkov photons 
are emitted at an angle with respect
to the particle track given by $\cos \, \theta_c = \frac{1}{\beta n_p}$.
The individual photons travel in the medium at the group velocity. 
Both the phase and group refractive indices depend on the
wavelength of the photons and has the effect of making the emission
angle and the speed of light wavelength dependent. 
This velocity of light has been measured using
a set of pulsed light sources
(LEDs emitting at different wavelengths) 
distributed throughout the detector illuminating the
PMTs through the water~\cite{velocity}.  
In special calibration data runs the emission
time and the position of the isotropic light flash,
as well as the arrival time and the position when the light reaches 
the PMTs are used to measure the velocity of light.
The refractive index has been measured 
at eight different wavelengths between 385~nm and 532~nm.  
This refractive index with its 
systematic errors are shown in Figure \ref{fig:showers} left.
Also shown is the parametric formula of the refractive index. 
The measurements are in agreement with the 
parametrization of the group refractive index.

\subsection{Measurement of Electromagnetic Showers along Muon Tracks}
The ANTARES detector measures mainly downward going muons. 
These muons are the decay products of cosmic ray 
collisions in the Earth's atmosphere.
Atmospheric muon data have been used for 
several analyses~\cite{4gev,anysotropy,fluxatmospheric}. 
In particular 
the collaboration investigated the sensitivity of 
the composition of cosmic rays through the
downward going muon flux~\cite{Hsu}. Several observational parameters
are combined to estimate the relative contribution of 
light and heavy cosmic rays. One of these parameters is the number of
electromagnetic showers along muon tracks.
 
Catastrophic energy losses appear occasionally, when a high energy muon 
($\sim 1$~TeV) traverses the water. 
These energy losses are characterized
by discrete bursts of Cerenkov light originating mostly from pair production
and bremsstrahlung (electromagnetic showers).
A shower identification algorithm~\cite{emshower,emshower1} 
is used to identify the excess of photons above 
the continuous baseline of photons emitted by
a minimum-ionizing muon. With this method downward going 
muons with energies up to 100 TeV have been analyzed. 

The muon event rate as a function of the number of identified
showers is plotted
in \mbox{Figure \ref{fig:showers}} right.
The distribution shows
the results for data and a Corsika based simulation.
As can be seen, about 5\% of the
selected muon tracks have at least one well identified shower.
Also shown is the systematic uncertainty for the simulation, 
where the largest systematic errors arises from uncertainties
on the PMT angular acceptance and absorption length.


\begin{figure}   
  \setlength{\unitlength}{1cm}      
   \centering
   \begin{picture}(18.0,6.0)
     \put(0.5, 0.0){\includegraphics[height=.27\textheight]{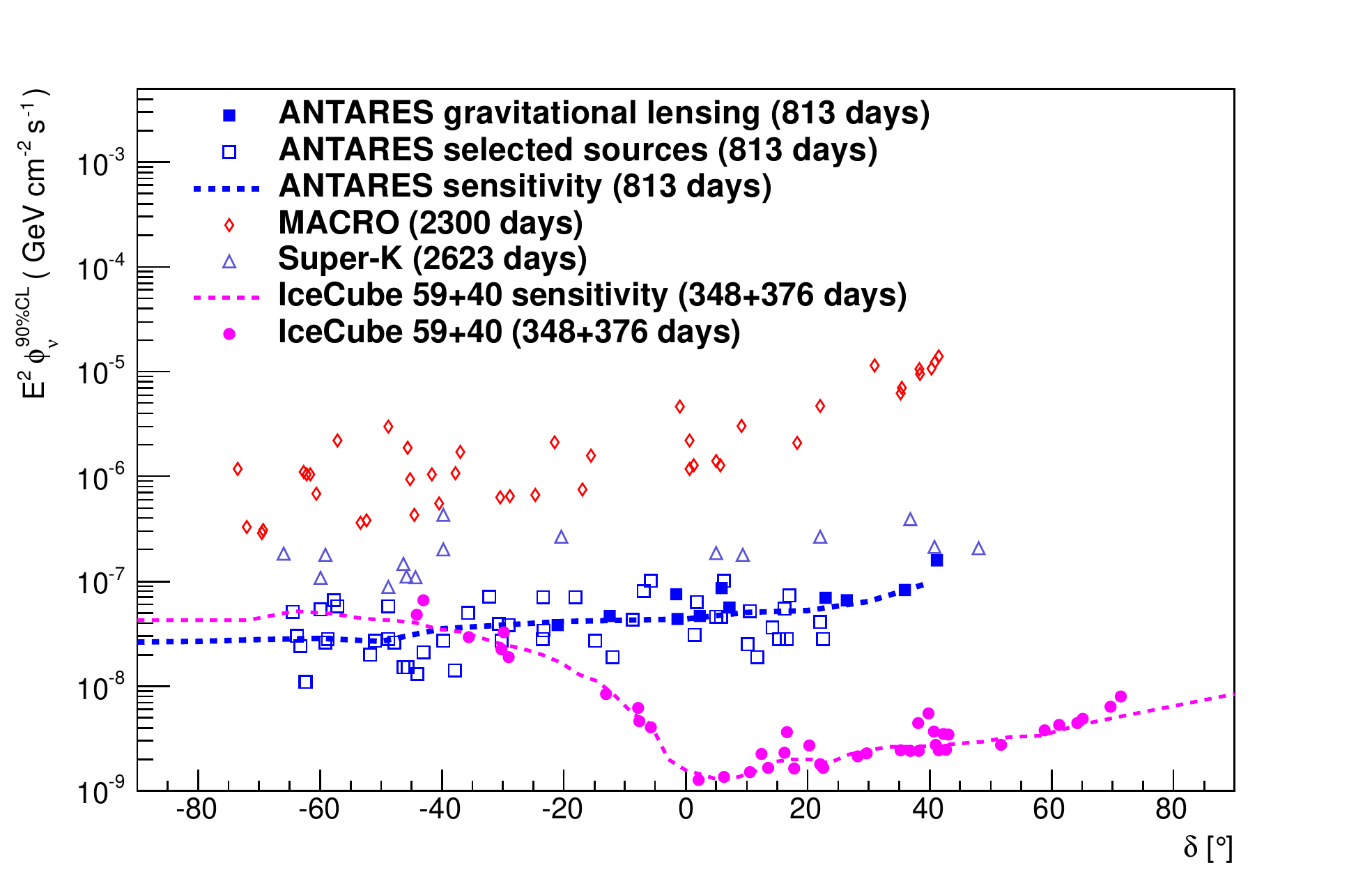}}
     \put(9.5, 0.0){\includegraphics[height=.27\textheight]{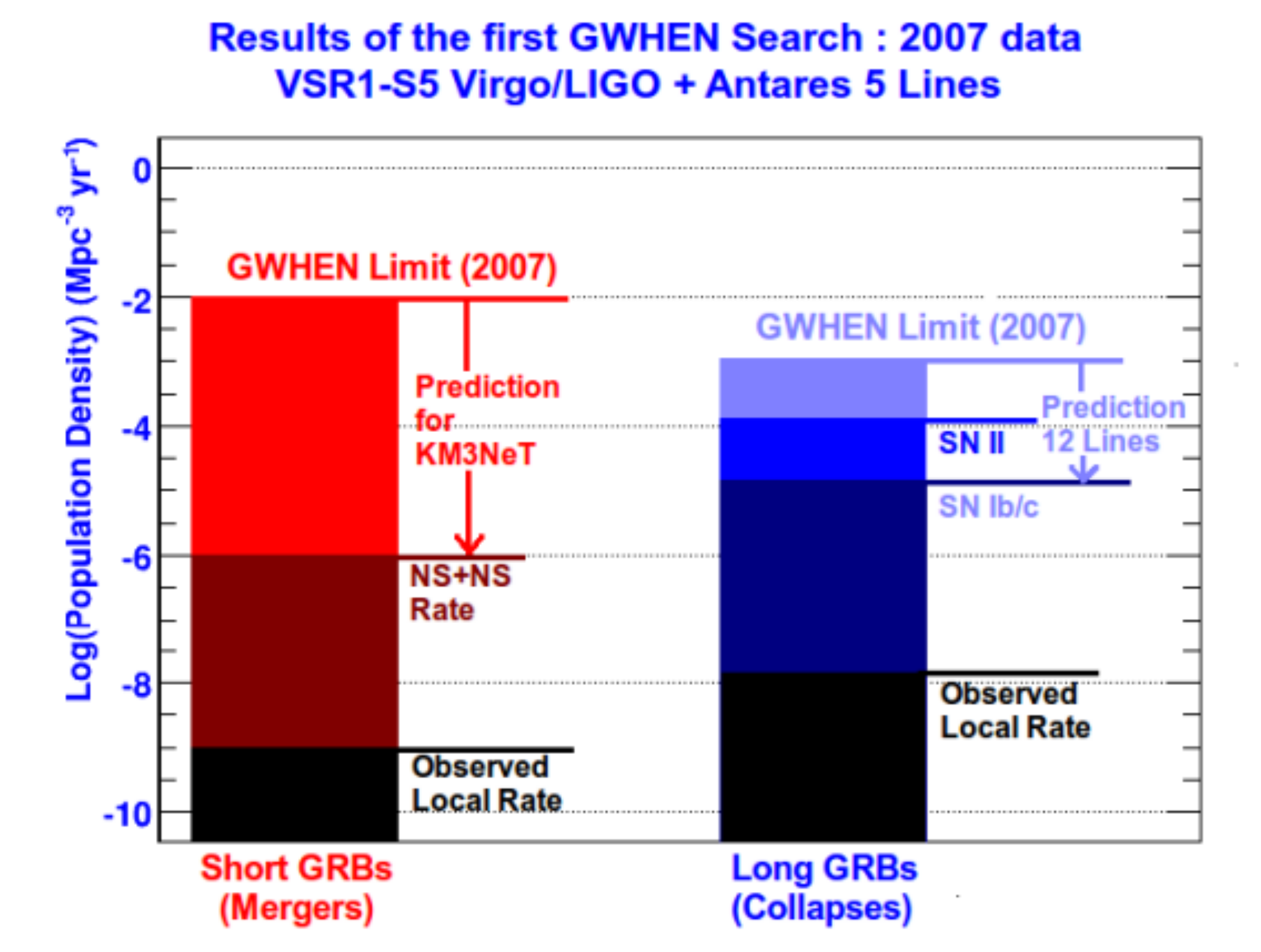}}
     \end{picture}
  \caption{\textit{Left: Upper limits (at 90\% C.L.) on the $E^{-2}$ neutrino flux from the 62 selected candidate sources as well as the sensitivity as function of the declination. Also shown are the results from other experiments. Right: Comparison of the gravitational waves (GW) and high energy neutrino (HEN) ANTARES limits of the local short and long GRBs rates, to the rate of neutron star - neutron star merger, and of type SN II and SN Ib/c rates. Also shown are potential reach of future analysis with ANTARES (12 Lines) and KM3NeT.}}
\label{fig:GW}
\end{figure}

\subsection{Search for Sources of Cosmic Neutrinos}

The collaboration has 
developed several strategies to search in its data 
for point-like cosmic neutrino sources~\cite{Pointsource2011,4yearpoint}, 
possibly in association with
other cosmic messengers such as gamma-rays~\cite{gammarayburst,gammaraypoint}, 
gravitational waves~\cite{gravitationalwave} or gravitational lenses.
Clustering of neutrino arrival directions 
can provide hints for their astrophysical origin.
In the search of cosmic neutrino point sources, upward going 
events have been selected in order to reject atmospheric muons. 
Most of the events are atmospheric muon neutrinos 
which constitute an irreducible diffuse
background for cosmic neutrino searches.
The 2007-2010 data contain around 3000 neutrino candidates with a predicted
atmospheric muon neutrino purity of around 85\%. The estimated angular
resolution is 0.46$\pm$0.10 degrees. 
The selection criteria are optimized 
to search for $E^{-2}$ neutrino flux from point-like 
astrophysical sources, following two different strategies: 
a full sky search and a search in the direction of 
particularly interesting neutrino candidate sources. 
The selection of 
these sources is either based on the intensity of 
their gamma-ray emission as observed by Fermi~\cite{Fermi} 
and HESS~\cite{Hess} or
based on strong gravitational lensed sources with large magnification.
The motivation to select lensed sources is that 
neutrino fluxes as photon fluxes 
can be enhanced by the gravitational
lensing effect, which could allow to observe 
sources otherwise below the detection
threshold. 

The cosmic point source search has 
been performed using an unbinned maximum likelihood
method~\cite{4yearpoint}.  
This method uses
the information of the event direction and, since the cosmic sources are
expected to have a much harder spectra than atmospheric neutrinos,
the number of hits produced by the track. 
For each source, the position of the cluster is fixed at
the direction of the source and the likelihood function is maximized
with respect to the number of signal events. 
In the absence of a significant excess of
neutrinos above the expected background, an upper limit on the
neutrino flux is calculated.
A full sky point source search based on the above mentioned algorithm
has not revealed a significant excess for any direction. 
The most significant cluster of events 
in the full sky search, with a post-trial 
$p$-value of 2.6\%, which is equivalent to $2.2\sigma$, corresponds to the 
location of $(\alpha,\delta)=(-46.5^{\circ},65.0^{\circ})$.  
No significant excess has been found neither in
the dedicated search from the list
of 11 lensed and 51 gamma-rays 
selected  neutrino source candidates.
The obtained neutrino flux limits of 
these selected directions are plotted as function of declination in
\mbox{Figure \ref{fig:GW}} left,
where for comparison the limits set by other neutrino experiments
are also shown.

\subsection{Search for Coincidence of Neutrinos and Gravitational Waves}
Both neutrinos and gravitational waves are cosmic messengers
that can escape from the core of the sources and travel over large distances through magnetic fields and matter without being altered. They could give important information about the processes taking place in the core of production sites and they could also reveal the existence of sources opaque to hadrons and photons such as failed GRBs. 
A first joint gravitational waves and neutrino search was performed using data taken with ANTARES and the gravitational waves detectors VIRGO and LIGO using 2007 data~\cite{gravitationalwave}. The strategy consists in an event-per-event search for  gravitational waves  signal correlation in space and time with a given high-energy neutrino event considered as an external trigger. No coincident event was observed, which allowed to place upper limits on the volume and density of joint gravitational waves and neutrino emitters. 
The gravitational waves horizon has been estimated to be $\sim$ 10 Mpc for mergers, and $\sim$ 20 Mpc for collapses.
The density limit ranges from  $10^{-2} Mpc^{-3} \times yr^{-1}$ for 
short GRB-like signals to   $10^{-3} Mpc^{-3} \times yr^{-1}$ for long GRB-like emission.
These density limits are presented in Figure \ref{fig:GW} right and are compared to other objects of interest.

\begin{figure}   
  \setlength{\unitlength}{1cm}      
   \centering
   \begin{picture}(18.0,5.8)
     \put(0.0, 0.0){\includegraphics[height=.24\textheight]{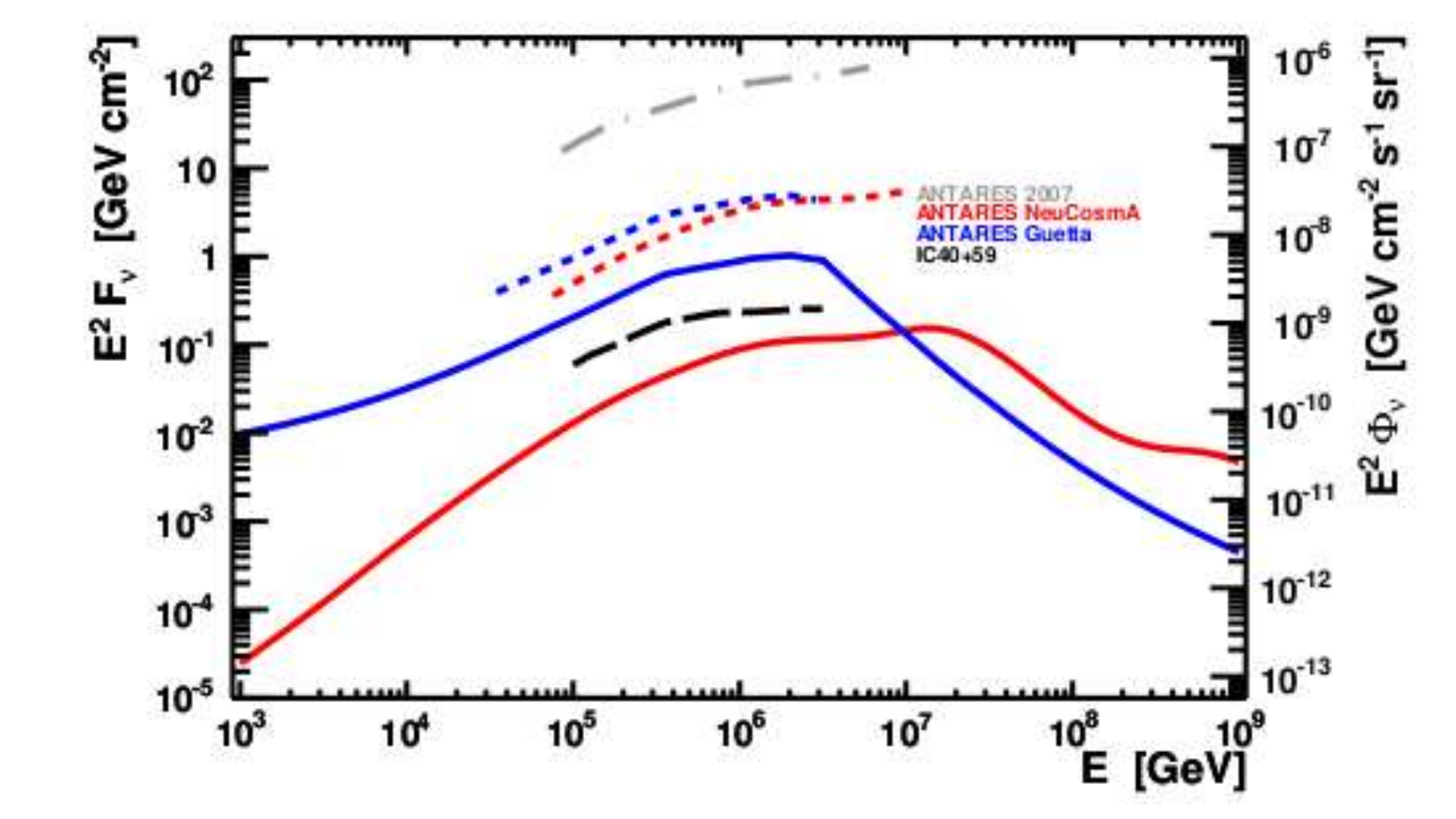}}
     \put(9.5, 0.0){\includegraphics[height=.24\textheight]{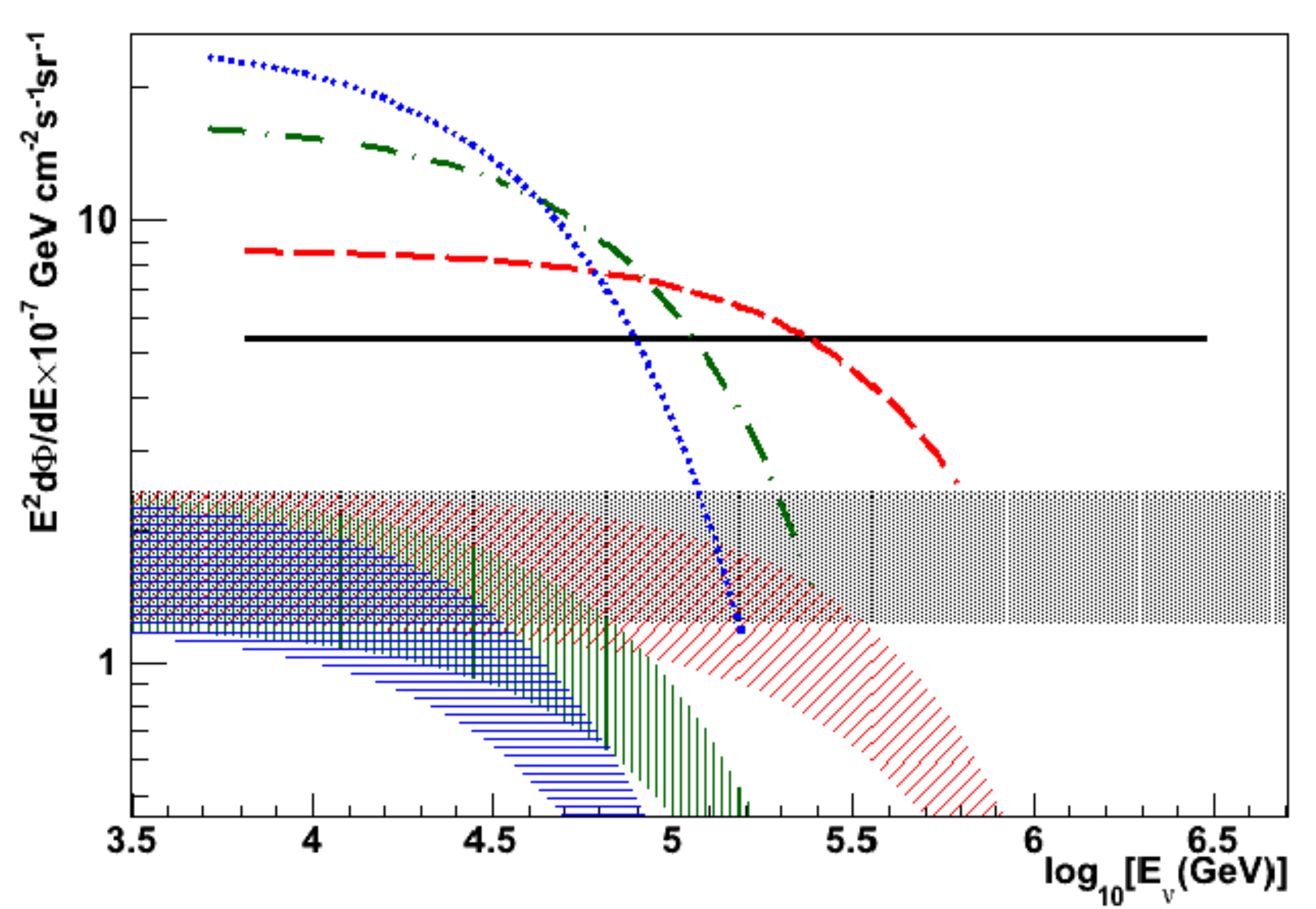}}
     \end{picture}
  \caption{\textit{Left: Comparison of 90\% C.L. experimental limits~\cite{icecubegrb,antgrb2,antgrb1} (dashed) with expected fluxes (solid) for the summed $\nu_{\mu}+\bar{\nu_{\mu}}$ spectra of the 296 GRBs as obtained from the NeuCosmA numerical simulation~\cite{neucosm} (red) or from the Guetta model~\cite{guetta} (blue). Right: Upper limits (at 90\% C.L.) on the $E^{-2}$ neutrino flux from the Fermi Bubbles for different cutoffs: no cutoff (straight black), 500 TeV (red), 100 TeV (green), 50 TeV (blue) are shown together with theoretical predictions with its uncertainties in case of pure hadronic model (colored regions). }}
\label{fig:FB}
\end{figure}

\subsection{Search for Neutrinos from Gamma Ray Bursts}
Another possible way to discover cosmic neutrinos is to observe neutrino events in coincidence
in direction and time with gamma ray bursts. Gamma ray bursts are intense flashes of gamma rays, 
which result from a highly relativistic jet formed during the collapse of a massive star such as supernova events.  
Two searches for a neutrino flux in coincidence with GRB have been made.
The first search selected 40 GRB that occurred in 2007~\cite{antgrb1}. The second search
is based on 2008-2011 data with 296 GRBs, representing a total equivalent live time of 6.55 hours.
In both cases, zero events were found in correlation with the photon emission of the GRBs. 
Figure~\ref{fig:FB} left shows the 
upper limits on the total flux for a fully numerical neutrino model which include Monte Carlo simulation~\cite{neucosm} and an analytical model~\cite{guetta}. 

\subsection{Search for Neutrinos from Fermi Bubbles}
The Fermi Satellite has revealed an excess of gamma-rays
in an extended pair of bubbles above and below our Galactic Center.
These so called Fermi Bubbles (FBs) cover about 0.8 sr of the sky,
have sharp edges, are relative constant in intensity and have a flat $E^{-2}$ 
spectrum between 1 and 100 GeV. 
It has been proposed that FBs are seen due
to cosmic ray interactions with the interstellar medium, which produce pions~\cite{Fermib}. 
In this scenario gamma rays and high-energy neutrino emission are expected with
a similar flux from the pion decays.

ANTARES has an excellent visibility to the FBs and therefore
a dedicated search for an excess of neutrinos in the
region of FB has been performed~\cite{FBubbles}. This analysis compares  
the averaged rate of observed neutrino events in the three FBs regions to 
that observed excluding the FB region. One such off source FB
region is equivalent in size and has in
average the same detector efficiency as the FB region.
The analyzed 2008-2011 data reveal 16 neutrino events inside the
FB region. Estimations from outside the FBs regions predict 
11 neutrino events. These results are compatible with no signal and
limits are placed on the fluxes of neutrinos 
for various assumptions on the energy 
cutoff at the source. Figure \ref{fig:FB} right shows the upper limits 
and compares it to expected signal 
for optimistic models~\cite{Fermib}. It can be seen 
that the calculated upper limits
are within a factor 3 above the expected signal.

\begin{figure}   
  \setlength{\unitlength}{1cm}      
   \centering
   \begin{picture}(18.0,5.8)
     \put(-0.0, 0.0){\includegraphics[height=.26\textheight]{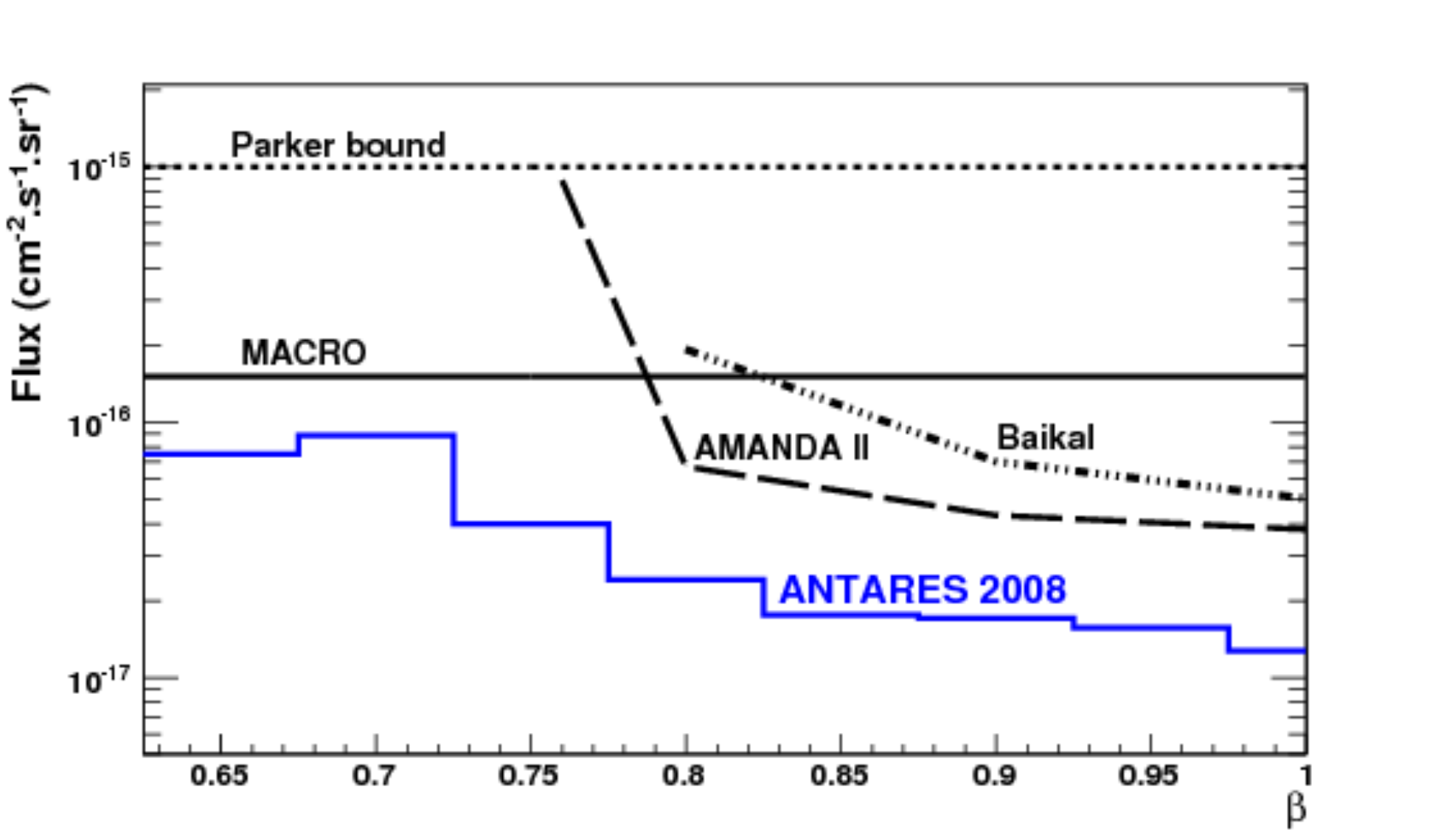}}
     \put(9.5, 0.0){\includegraphics[height=.24\textheight]{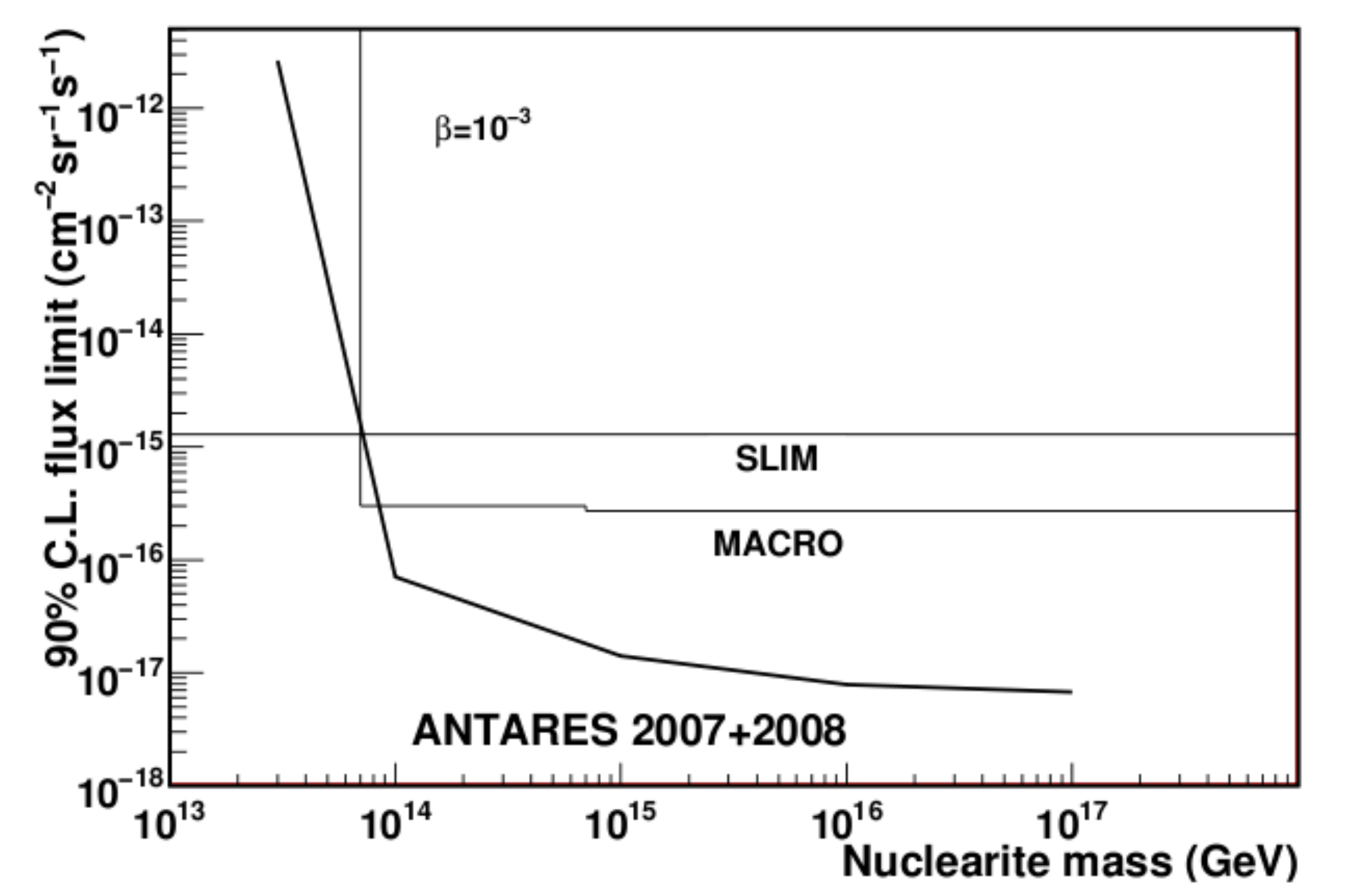}}
     \end{picture}
  \caption{\textit{Left: The ANTARES 90\% C.L. upper limit on the upgoing magnetic monopole flux as a function of the monopole velocity $\beta$. Also shown are the theoretical Parker bound~\cite{Park}, the published upper limits obtained by MACRO~\cite{macromm} for an isotropic flux of monopoles as well as the upper limits from Baikal~\cite{baikalmm} and AMANDA~\cite{amandamm} for upgoing monopoles. Right: The ANTARES 90\% C.L. upper limits on a downgoing flux of nuclearites as a function of the nuclearite mass, compared to the limits reported by MACRO~\cite{macronuc} and SLIM~\cite{slimnuc}.}}
\label{fig:monopol}
\end{figure}

\subsection{Search for Relativistic Magnetic Monopoles}
The existence of magnetic monopoles is a generic prediction of grand unification theories and are expected to have been produced in the early Universe. The mass predicted for such magnetic monopoles can range from $10^{4}$~GeV to $10^{20}$~GeV depending on the specific model~\cite{monopol}. Magnetic monopoles are one of the few predictions of grand unification theories that can be studied in the present environment.
The detection of monopoles relies on the large amount of light emitted compared to that from muons. Cherenkov light emission of a 
monopole exceeds that of a single charged minimum ionizing particle by  $O(10^{4})$. The search for upgoing relativistic magnetic monopoles was performed with 116 days live time of ANTARES 2007-2008 data~\cite{antaresmonopol}. One event was observed, consistent with the expected atmospheric neutrino and muon background. The derived limits on the
upgoing magnetic monopole flux for monopoles with velocity $\beta>0.625$ are shown in Figure \ref{fig:monopol} left.    

\subsection{Search for Slowly Moving Nuclearites} 
Nuclearites are hypothetical massive particles assumed to be stable lumps of up, down and strange quarks in nearly equal proportions. They could be present in the cosmic rays relics of the early Universe. 
The nuclerite detection in neutrino telescopes is possible through the blackbody radiation emitted by the
expanding thermal shock wave along their path~\cite{nuclear}.  
A search was performed for downgoing 
slowly moving nuclearites ($\beta \sim 10^{-3}$) with data collected in 2007 and 2008~\cite{nuclearites}. 
Only nuclearites with masses larger than a few $10^{13}$ GeV produce enough light to be detected within the detector. 
A dedicated search strategy found no significant excess of nuclearite events.
The upper limits on the flux of downgoing nuclearites are shown in Figure \ref{fig:monopol} right for the mass range $\sim 10^{13}-10^{17}$ GeV.

\section{Conclusion}
ANTARES has been taking data since the first lines were deployed in 2006
and is foreseen to take data at least until the end of 2016.
With these data a broad physics program is underway 
producing competitive results. In particular the neutrino telescope has continuously monitored the Southern Sky looking for TeV neutrino sources, but unfortunately ANTARES has still not seen 
any cosmic neutrinos.  The next generation multi-km$^{3}$ neutrino telescope KM3NeT has started to be built in the Mediterranean Sea. This will complement the field of view of the IceCube detector at the South Pole. Neutrino telescopes 
are starting to open up a new window in the sky exploring new territory
and they will hopefully reveal new unknown phenomena and 
help answer open questions.




\begin{theacknowledgments}
I gratefully acknowledge the support of the JAE-Doc postdoctoral program of CSIC.
This work has also been supported by the following
Spanish projects: FPA2009-13983-C02-01, MultiDark Consolider CSD2009-00064, ACI2009-1020 of MICINN and Prometeo/2009/026 of Generalitat Valenciana.
\end{theacknowledgments}


\begin{thebibliography}{9}


\bibitem  
[A. Achterberg  et al., \emph{Astropart. Phys.}  26 (2006) 155.]{icecube}


\bibitem{Ant1} 
[M. Ageron et al.,  \emph{Nucl. Instrum. Methods} A 656 (2011) 11.]{Ant1}



\bibitem{4gev} 
J.A. Aguilar et al., \emph{Astropart. Phys.} 33 (2010) 86.

\bibitem{diffusepaper} 
J.A. Aguilar  et al.,  \emph{Phys. Lett.} B696 (2011) 16.  

\bibitem{Dmatter} 
S. Adrian-Martinez  et al., \emph{arXiv:1302.6516} (2013).


\bibitem{timecalib}   
J.A. Aguilar et al.,  \emph{Astropart. Phys.} 34 (2011) 539.

\bibitem{acoustic}   
J.A. Aguilar et al., \emph{Nucl. Instrum. Methods} A626 (2011) 128.





\bibitem{amandaflux}
R. Abbasi et al.,  \emph{Astropart. Phys.} 34 (2010) 48.

\bibitem{icecubeflux}
R. Abbasi et al.,  \emph{Phys. Rev.} D83 (2011) 012001.

\bibitem{barr}
G.~D. Barr et al., \emph{Phys. Rev.} D 70 (2004) 023006.


\bibitem{Martin}
A.D. Martin et al., \emph{Acta Phys. Polon} B34 (2003) 3237.

\bibitem{Enberg}
E. Enberg et al.,  \emph{Phys. Rev.} D78 (2008) 034005.




\bibitem{oscillation} 
S. Adrian-Martinez et al., \emph{Phys. Lett.} B714 (2012) 224.

\bibitem{minos} 
P. Adamson et al., \emph{Phys. Rev. Lett.} 101 (2008) 131802.

\bibitem{fastalgo}  
J.A. Aguilar et al., \emph{Astropart. Phys.} 34 (2011) 652.




\bibitem{velocity} 
S. Adrian-Martinez et al., \emph{Astropart. Phys.} 35 (2012) 552.



\bibitem{anysotropy} 
S. Mangano et al., \emph{arXiv:} 0908.0858 (2009).

\bibitem{fluxatmospheric} 
J.A. Aguilar et al., \emph{Astropart. Phys.} 34 (2010) 179.


\bibitem{Hsu} 
C. Hsu et al., ICRC, HE2.3, 0679 (2011).

\bibitem{emshower} 
J.A. Aguilar et al., \emph{Nucl. Instrum. Methods} A675 (2012) 56.

\bibitem{emshower1} 
S. Mangano  et al., \emph{Nucl. Instrum. Methods} A581 (2007) 695.




\bibitem{Pointsource2011} 
S. Adrian-Martinez et al.,   \emph{Astrophys. J. Lett.} 743 (2011)  L14.  

\bibitem{4yearpoint} 
S. Adrian-Martinez et al.,  \emph{Astrophys. J.} 760 (2012)  53. 

\bibitem{gammarayburst} 
S. Adrian-Martinez  et al.,  \emph{JCAP} 1303 (2013) 006.

\bibitem{gammaraypoint} 
S. Adrian-Martinez  et al.,   \emph{Astropart. Phys.} 36 (2013) 634.

\bibitem{gravitationalwave} 
S. Adrian-Martinez et al.,    \emph{JCAP} 1306 (2013) 008.




\bibitem{Fermi}  
W.B. Atwood et al.,   \emph{Astrophys. J.} 697 (2009) 1071.

\bibitem{Hess}  
K. Bernloehr  et al.,   \emph{Astropart. Phys.} 20 (2003) 111.



\bibitem{icecubegrb}
R. Abbasi et al., \emph{Nature} 484 (2012) 351.

\bibitem{antgrb2}
S. Adrian-Martinez et al.,  \emph{arXiv:} 1307.0304.

\bibitem{antgrb1}
S. Adrian-Martinez et al.,  \emph{JCAP} 1303 (2013) 006.

\bibitem{neucosm}
S. Hummer et al., \emph{Phys. Rev. Lett.} 108 (2012) 351.

\bibitem{guetta}
D. Guetta et al., \emph{Astropart. Phys.} 20 (2004) 429.




\bibitem{Fermib} 
R. Crocker et al., \emph{Phys.Rev.Lett.}  106 (2011) 101102.

\bibitem{FBubbles}  
S. Adrian-Martinez et al., \emph{arXiv:1308.5260}  (2013).



\bibitem{monopol}
J.~Preskill, \emph{Ann. Rev. Nucl. Part. Sci.} 34 (1984) 461.

\bibitem{antaresmonopol}  
S. Adrian-Martinez et al,  \emph{Astropart. Phys.} 35 (2012) 634.

\bibitem{Park}
E.~N.~Parker, \emph{Astrophys. J.} 160 (1970) 383.

\bibitem{macromm}
M.~Ambrosio et al., MACRO Coll. \emph{Eur. Phys. J.} C25 (2002) 511.

\bibitem{baikalmm}
V.~Aynutdinov et al., BAIKAL Coll. \emph{Astropart. Phys.} 29 (2008) 366.

\bibitem{amandamm}
R.~Abbasi et al., IceCube Coll. \emph{Eur. Phys. J.} C69 (2010) 361.




\bibitem{nuclear}
A.~De Rujula, S.~L. Glahshow, \emph{Nature} 312 (1984) 734.

\bibitem{nuclearites}  
G.E. Pavalas et al., \emph{Rom. Rep. Phys.} 64 (2012) 325.

\bibitem{macronuc}
M.~Ambrosio et al.,  \emph{Eur. Phys. J.} C13 (2000) 453.

\bibitem{slimnuc} 
S.Cecchini et al., \emph{Eur. Phys. J.} C57 (2008) 525.



\end{thebibliography}
\end{document}